\documentclass{emulateapj}
\usepackage{epsfig}
\usepackage{graphicx}
\usepackage{graphics}
\usepackage{color}

\shorttitle{Mass--Radius Response in Dynamical Mass Loss}
\shortauthors{Woods, Ivanova}

\begin{document}

\title{Can We Trust Models for Adiabatic Mass Loss?}

\author{T.E.\ Woods}
\affil{University of Alberta, Dept.\ of Physics, 11322-89 Ave, Edmonton, AB, T6G 2E7, Canada}

\author{N.\ Ivanova}
\affil{University of Alberta, Dept.\ of Physics, 11322-89 Ave, Edmonton, AB, T6G 2E7, Canada}

\begin{abstract}

In interacting binaries, comparison of a donor star's radial response to mass loss with the response of its Roche radius determines whether mass loss persists and, if so, determines the timescale and stability of the ensuing evolutionary phase. For giants with deep convective envelopes, the canonical description holds that once mass transfer begins it typically proceeds catastrophically on the dynamical timescale, as the star cannot lose sufficient heat in order to avoid expansion. However, we demonstrate that the {\bf local} thermal timescale of the envelope's superadiabatic outer surface layer remains comparable to that of mass loss in most cases of ``dynamical" mass loss. We argue therefore that if mass loss proceeds on a timescale longer than this, then even a deep convective envelope will not dramatically expand, as the surface layer will have time to relax thermally and reconstitute itself. We demonstrate that in general the polytropic approximation gives much too strict a criterion for stability, and discuss the dependence of the donor's response on its radius in addition to its core mass. In general, we find that the effective response of the donor on rapid timescales cannot be determined accurately without detailed evolutionary calculations.  

\end{abstract}

\section{Introduction}

For evolved giant branch stars in binary systems the canonical view holds that, should the giant fill its Roche lobe (RL), the ensuing mass loss (ML) would in most circumstances be dynamically unstable. This is due to the donor's deep convective envelope, as it is generally held that a convective envelope invariably expands upon ML \citep{Pac72}. At the same time the orbit could either shrink, or expand slower than is required to keep the expanding donor inside its RL. Run-away ML follows and would then lead to an event typically referred to as common envelope (CE) evolution, as originally conceived by \cite{Pac76} to explain the formation of cataclysmic variables.

The first quantitative limits on the stability of ML in such systems were found by \cite{Hjellming87}, by modelling the donor star as a condensed polytrope -- specifically, an $n = 3/2$ polytrope with a point mass at the center to model the core, implying an isentropic profile throughout the envelope. The rapid expansion of these models in response to ML further confirmed the susceptibility of giant donors to dynamically unstable runaway events.

The calculations were performed in the so called adiabatic regime, which in subsequent studies became a standard approach to analyze dynamical instability with respect to mass transfer (MT) in binaries. It is important here to clarify what is assumed under the ``adiabatic" approximation:  a) a (stellar or polytropic) model is always in hydrostatic equilibrium (HE); b) a model keeps its initial entropy profile as a function of mass  \citep{Hjellming87, GeWebbink08, Ge_etal2010, GeWebbinkHanChen2010}. This means that a characteristic timescale $\tau_{\rm ad}$ is assumed, which is always shorter than the thermal timescale of the star $\tau_{\rm th}$, but longer than the dynamical timescale $\tau_{\rm dyn}$. Therefore, the obtained model is valid only so long as the implied $\tau _{\rm{ML}}$ satisfies this condition in all layers. Models with reduced mass obtained as the result of adiabatic ML calculations ought to be driven out of thermal equilibrium. 

Consequently, the accuracy of the adiabatic model rests on the assumption that the outermost layer of the giant, in which convection becomes highly inefficient, is negligible in determining MT stability. This can be interpreted as assuming that such a layer is very quickly stripped at the onset of dramatic ML, with the entropy profile through all remaining layers unable to adjust in order to restore thermal equilibrium. 

In this {\it Letter} we argue that in this case the shorter thermal timescale of the outer envelope would allow it to relax at a rate comparable to or even faster than both the expected ML or $\tau_{\rm dyn}$, implying that an adiabatic approach (either by considering a polytropic, or even a realistic stellar structure) would model the reaction of the envelope rather poorly. We examine the effect of accounting for the thermal relaxation of the outer envelope, and find that in many cases the envelope will in fact continue to contract with ML despite the deep convective envelope.

\section{Response of the donor to ML}

The formalism for the response to ML of a giant branch donor (modelled as a condensed polytrope) is well-established \citep{Hjellming87, Soberman97}. Here we briefly review the canonical treatment for stability relations related to the stellar mass-radius response. In such analysis, stability is determined from considering small perturbative variations about the equilibrium state, in which the donor just fills its RL, giving: 

\begin{center}
\begin{equation}\label{zeta}
\Delta \zeta = \frac{M}{R}\frac{\delta \Delta R}{\delta M}
\end{equation}
\end{center}

\noindent where $\zeta$ is the response exponent from the mass-radius relation ($R \propto M^{\zeta}$), $M$ and $R$ are the donor mass and radius, and $\Delta R = R - R_{\rm RL}$ is the difference between the RL radius $R_{\rm RL}$ and the donor radius. Depending then on the strength of the donor radius' response, ML may proceed on one of three timescales: dynamical, thermal, or nuclear ($\tau_{\rm nuc}$). 

For adiabatic ML, a giant star with an isentropic envelope is modelled as a condensed polytrope; using this \cite{Hjellming87} found that the adiabatic mass--radius exponent of a giant branch star follows to good approximation:

\begin{center}
\begin{equation}
\zeta_{\mathrm{ad}} \equiv \left ( \frac{d\log R}{d \log M} \right )_{\rm ad} = \frac{2}{3} \ \frac{m_\mathrm{c}}{1-m_\mathrm{c}} - \frac{1}{3} 
\end{equation}
\end{center}

\noindent where $m_{\mathrm{c}}$ is the core-mass fraction of the donor. Comparison of $\zeta_{\rm ad}$ with the radius-mass exponent of the RL ($\zeta_{\rm RL}$) then provides the condition on the binary mass ratio, above which it is believed dynamically unstable MT occurs.

As was mentioned in \S1, studies of adiabatic ML thus far assume a state of HE in constructing their stellar models, implicitly taking adiabatic ML to be at a timescale shorter then the thermal timescale of the donor but longer then the characteristic dynamical timescale. In modelling this process outer mass shells are removed while maintaining a constant entropy profile with respect to the mass coordinate. However if the ML is modeled to be adiabatic as described above, such that $\tau_{\rm ML}\equiv\tau_{\rm ad} > \tau_{\rm dyn}$, then the modelled stellar reaction,  by definition, is {\bf not dynamical}, but thermal. It is inherently misleading then, in this type of analysis, to refer to the process as {\it dynamical ML}, as is typically done.

In real stars, mass removal from the outer layer of the donor will perturb the HE of the star, to which it responds on its dynamical timescale. Indeed, as long as the ML proceeds on a timescale $\tau_{\rm ML}>\tau_{\rm dyn}$, the star and its envelope remain in HE.  In this case, the reaction of the radius of the star is determined by the reaction on the next, longer timescale, presumably thermal, when the internal heat is redistributed in the envelope. Deep within the envelope, convection is adiabatic; this is very efficient and maintains the flat entropy profile (also, its value will be constant if $\tau_{\rm ML}<\tau_{\rm th}$). However, in the upper layers of the envelope the density drops off dramatically; convection here is much less efficient and radiative cooling wins out; the outer reaches of the giant become superadiabatic ($\nabla>\nabla _{\mathrm{ad}}$). The timescale to redistribute heat in this upper layer could be significantly different from that estimated for deep regions based on adiabatic convection. It is the relaxation of the outer layers which drives MT, giving the relevant timescale at high $\dot M$ \citep{Pod02}.  
 
It is important to emphasize that if the ML occurs either on the dynamical timescale,  or on a timescale longer than the time necessary for the outer layers to obtain their thermal equilibrium, the adiabatic approach is intrinsically invalid.

\begin{figure}
\includegraphics[height=.35\textheight]{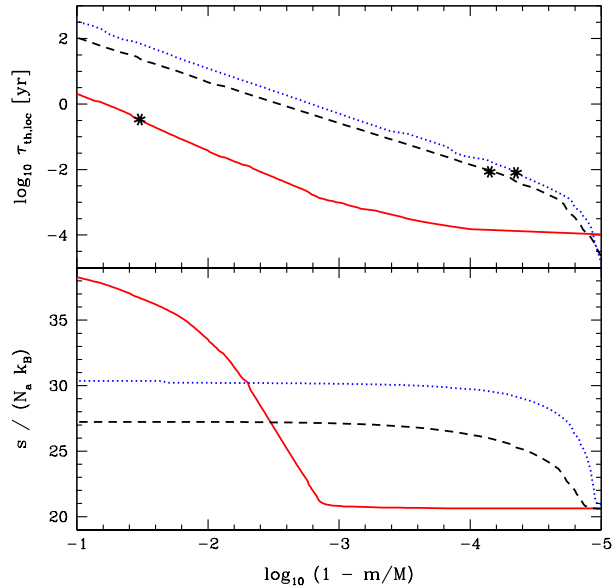}
\caption{Entropy (lower panel) and local thermal timescale (upper) in the outer 10\% mass of a $1~M_\odot$ ($R=30~R_\odot$), $5~M_\odot$ ($R=50~ R_\odot$) and $20~M_\odot$ ($R=950~R_\odot$) giant (black dashed, blue dotted and red solid lines respectively). Here, thermal timescale was calculated using stellar models as  $\tau_{\rm{th,loc}}=\int_m^{M} u(m) dm / L$, where $u$ is the specific internal energy. Asterisks indicate the locations to the right of (above) which $\tau_{\rm{th,loc}} < \tau_{\rm{dyn}}$.}\label{profile}
\end{figure}

Let us now consider realistic stellar models. Within real giant branch stars, the entropy profile is marked by a steep drop-off in the outermost superadiabatic ``surface" of the envelope (see Figure~\ref{profile}). In prior work concerning adiabatic ML it is generally assumed that this layer is quickly removed, exposing the convective ($dS/dM = 0$) layer beneath. We note that it is only in this case that the donor's structure is well-approximated by a condensed polytropic model, in which case one expects dramatic expansion with ML \citep{Ge_etal2010}. It can be seen as well that since in massive giants the superadiabaticity plays a role for deep layers as well, the adiabatic approach that assumes a flat entropy profile is also invalid. Also, for  all giants, the local thermal timescale of the superadiabatic layer $\tau_{\rm th,loc}$ is {\it shorter} than the donor's dynamical timescale: during the thermal relaxation of this layer, assumption of HE is not valid a priori.

For real stars, it is not immediately clear that we can safely ignore the response of the surface of the envelope to ML. In principle, the rate of ML must surpass the thermal timescale of the superadiabatic outer envelope in order for this layer to be effectively removed, and the standard picture of adiabatic ML to take hold. Otherwise, this layer would have time to thermally relax, reconstituting itself faster than it can be torn away. The thermal timescale of the superadiabatic outer shell, roughly, is $\sim GMm_{\mathrm{shell}}/RL$ (here $m_{\rm shell}$ is the mass of the shell that has a non-isentropic profile and $L$ is the star's luminosity) and the MT timescale as $\tau _{\mathrm{ML}} = M/\dot M$. We then have a minimum ML rate that would allow for this outer surface of the envelope to be stripped away:

\begin{equation}
\dot M_{\mathrm{crit}} \geq 3.2  \frac{M_{\odot}}{\mathrm{yr}} \times 
\left ( \frac{R}{100 R_{\odot}}\right )  \left (\frac{L}{100 L_{\odot}}\right )
\left (\frac{10^{-4} M_{\odot}}{m_{\mathrm{shell}}}\right )
\label{criterion}
\end{equation}

\noindent where with substantially lower MT rates any response of the donor to ML would be poorly modelled using a polytropic model. With this limit, a 5 $M_\odot$ giant with $m_{\mathrm{shell}} \sim 0.0001 M_{\odot}$, $L \sim 1000 L_{\odot}$, and $R \sim 50 R_{\odot}$ would need a minimum ML rate of $\dot M_{\mathrm{crit}} \sim 16 M_{\odot}/\mathrm{yr}$ in order to strip the superadiabatic surface layer. For ML rates less than this, we cannot assume the envelope's surface layer is removed.

To test the response in real stellar models, we performed simulations using two different stellar evolutionary codes. 
The first is a binary stellar-evolution code {\tt STARS/ev}, originally developed by Eggleton \citep{Egg71, Egg72, Egg73, Eggetal73} and then updated by many others, as detailed in \cite[][and references therein.]{Pols95,Glebbeek2008}. 
It is a non-Lagrangian code that only calculates stellar model in HE, but is capable of treating stars out of thermal equilibrium. MT is treated in a parametrized form using

\begin{equation}\label{mdot}
\dot M = C\cdot \ln\left(\frac{R}{R_{\rm{RL}}}\right) \cdot 10^{33} M_{\odot}\rm{yr}^{-1}
\end{equation}

\noindent where C is a free parameter with values ranging from 1 to $10^{4}$ (van der Sluys \& Glebbeek, private communication, see also \cite{NelEgg, EggEgg}). The second is a standard Henyey-type code, as derived from that originally developed in \cite{Kip67}; its current state and input physics are described in \cite{Iva03, Iva04}. This code is Lagrangian, and is capable of treating stars out of both hydrostatic and thermal equilibrium. Here, MT in a binary is calculated implicitly keeping the donor always within its RL.

\begin{figure}[ht]
\includegraphics[height=.25\textheight]{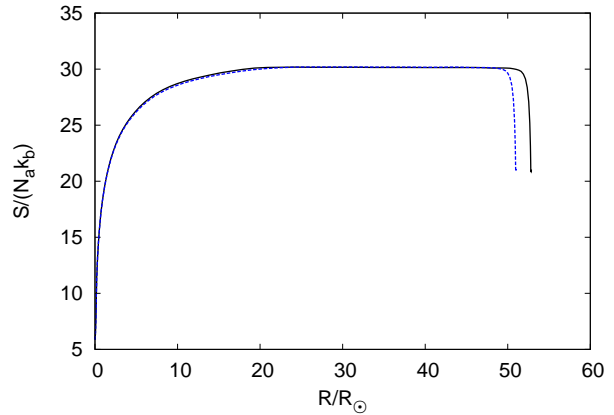}
\caption{Entropy profile of a 5$\mathrm{M}_{\odot}$ giant undergoing steady ML at a rate of $10^{-2}\mathrm{M}_{\odot}\mathrm{yr}^{-1}$ at the onset (black, solid), and after losing $\sim 0.6 M_{\odot}$ (blue, dashed).}\label{ML}
\end{figure}

Using this we can apply a constant rate of ML and observe the star's response. As an example, consider a $5M_{\odot}$ star on the giant branch with a radius of $\sim 50R_{\odot}$. Applying ML rate $\dot M=10^{-2}M_{\odot}\mathrm{yr}^{-1}$ (the canonical adiabatic ML regime for this star would be $\dot M \ga 10^{-4}~M_\odot {\rm yr}^{-1}$) we see that the radius shrinks with ML, despite the deep convective envelope of the donor. Note that this result is independent of the code itself, as we find the same effect using both codes mentioned above. The reason for this becomes clear in examining Figure~\ref{ML} -- this is because a superadiabatic outer layer remains present even after the giant has rapidly lost $~0.6M_{\odot}$.  The only difference we found is that with the second code $m_{\rm sh}$ is smaller, providing {\it greater} stability (see Equation~\ref{criterion}.) 

Rather than being negligible, understanding the response of this surface layer remains an important factor in determining the true response of a giant donor's radius to ML. In order to better understand the impact of this, we must examine how this impacts MT stability in real binaries. 

\section{Discussion}

A differential change in the donor radius can be expressed as:

\begin{equation}  
d \ln R_{d} = \left( \frac{\partial \ln R_{d}}{\partial t}\right)dt + \left(\frac{\partial \ln R_{d}}{\partial \ln M_{d}}\right)d\ln M_{d} \ ,
\end{equation}

\noindent where the second term describes the variation in the donor's radius with ML over short time intervals, implying constant entropy and composition, and gives the adiabatic response of the donor. 
The first term can be broken up into the normal evolution of the star, and the response of the radius during the star's adjustment on its thermal and dynamical timescales. The former is negligible on any timescale considered here. Allowing for the relaxation of the outer surface of the envelope, we then have an effective mass-radius exponent:

\begin{equation}\label{zeta_eff}
\zeta _{\mathrm{eff}} = \frac{d\ln R_{\mathrm{d}}}{d\ln M_{\mathrm{d}}} =   \frac{d\ln R_{\mathrm{d}}}{\partial t} \frac{M_{\mathrm{d}}}{\dot M_{\mathrm{d}}} + \zeta _{\mathrm{ad}}
\end{equation}

\begin{figure}[ht]
\includegraphics[height=.37\textheight]{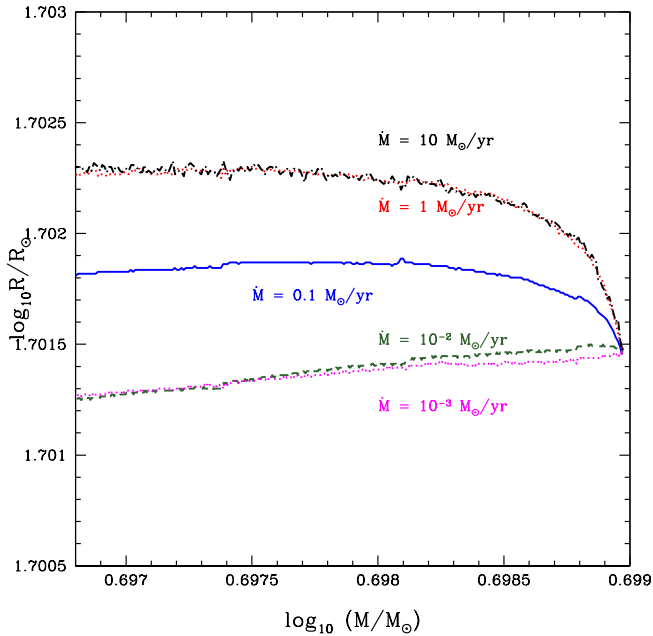}
\caption{Mass-radius response for varying ML rates applied to an initially 5 $M_{\odot}$ giant with radius 50 $R_{\odot}$.}\label{zeta_5m}
\end{figure}

\noindent For $\tau _{\mathrm{th, loc}} \la \tau_{\mathrm{ad}}$, the thermal relaxation of the donor's superadiabatic surface can dramatically impact the effective response of the donor. However, calculations show that even for extremely rapid ML ($10M_{\odot}\mathrm{yr}^{-1}$) this outer layer is not fully removed. For MT rates near the critical rate estimated by eq.~\ref{criterion} the donor radius expands,  but not nearly to the dramatic extent we would expect in the polytropic approximation:  the {\it maximum} expansion is $\sim0.15\%$ of the initial radius (see Figure~\ref{zeta_5m}), while in the case where the superadiabatic layer is removed, it is expected to expand by $\sim 30\%$ 
\citep[see e.g. Figure~6 in ][]{Ge_etal2010} -- the relative radial expansion in MT calculations with real stllar models is {\it 200 times less!} Well below $\dot M_{\mathrm{crit}}$, although within what is typically considered the adiabatic regime, the donor contracts. 

The consequences for MT stability in a binary are substantial. For a $5M_{\odot}$ giant donor with $R=50R_{\odot}$ and a core mass of $m_{\rm c}=0.856 M_{\odot}$, the condensed polytropic approximation for the donor's response predicts unstable MT for any mass ratio  $q=M/M_{\rm comp} > 0.856$ ($M_{\rm comp}$ is the companion mass). 
However, with the binary stellar-evolution code {\tt STARS/ev}, 
the stable MT proceeds for mass ratios as high as $q=1.47$ (with C = 1), 
down to the nearly complete removal of the original envelope ($<1\%$) 
\footnote{We provide in this stability analysis the results of {\tt STARS/ev} as it gives 
{\it least} stability in binaries compared to the second code.}.

The values given above are not hard limits; using STARS/ev, it is difficult to decide in principle when any given run has a true run-away, and when the problem is numerical. In fact, for $10M_{\odot}$ at $300R_{\odot}$, we are unable to evolve the system through RL overflow even below $q_{\rm{HJ,crit}}$ without $\Delta R > 10\%$, which we assume leads to runaway \citep[see ][]{Pod02}. The parametrization given in eq. \ref{mdot} is crude at best, and in general a better understanding of the degree to which a donor may overflow its RL (and the resulting MT rate) are needed. We note that for the maximum value of C = 10000 typically used in STARS (Eggleton 2010, private communication), we find $q_{\rm{crit}}\approx 1.25$, although high values of C are known to cause the model to crash at the onset of MT (van der Sluys \& Glebbbeek 2010, private communication). 

\begin{figure}[ht]
\includegraphics[height=.27\textheight]{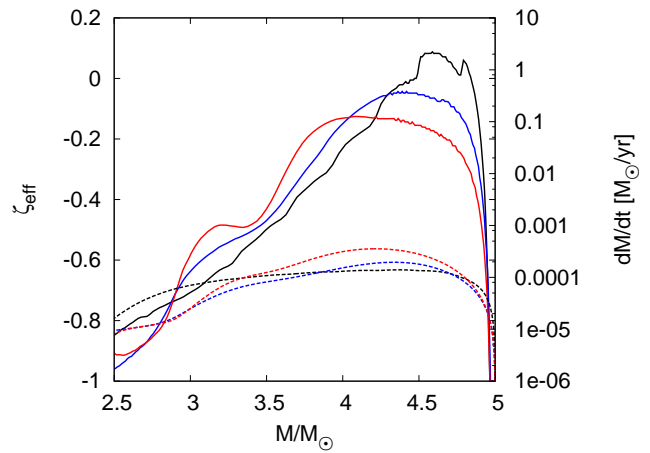}
\caption{Initial evolution of the effective radial response of the donor (solid lines) and $\dot M$ (dashed lines) during ML, for a 5$M_{\odot}$+5$M_{\odot}$ binary with a donor radius of 50$R_{\odot}$ (black, highest peak), 68$R_{\odot}$ (blue, middle) and 85$R{\odot}$ (red, lower) with a point mass companion.  }\label{response}
\label{zeta_fig}
\end{figure}

Continuing as above, we find a much broader range of stability against the onset of a dynamical run-away than previously thought, as demonstrated in Table~\ref{table}. In these simulations, we assume MT proceeds in a stable manner if the program terminates with either all of the original envelope being removed, or at least reaching the end of the first episode of MT \citep[for $2.0M_{\odot} \lesssim M_{\rm{d}} \lesssim 15M_{\odot}$, intermediate mass case B evolution, see ][]{InteractingBinaries}.  
We note as well that under the polytropic approximation a binary with $q=1$ could only avoid a dynamical instability for $m_{\rm{c}}>0.4567$ \citep{Hjellming87}; this condition does not appear in our calculations (see Figure~\ref{zeta_fig}, Table~1).

Defining $\zeta_{\rm eff}$ from simulations is non-trivial (see Figure~\ref{zeta_fig}):
after an initially weaker expansion than anticipated, $\zeta _{\mathrm{eff}}$ quickly rises before levelling off to a slow steady rise through the loss of the first $10\%$ of the donor's mass. In Table~\ref{table}, $\zeta_{\rm eff}$
is an averaged value of $\zeta$ for this phase, during which the binary evolves on the quasi-thermal timescale (having reached the peak of the $\tau_{\rm{th}}$ phase of case B MT, with $\dot M\sim 10^{-3}M_{\odot}/\rm{yr}$). 
We find as well that $\zeta_{\rm eff}$ decreases as $q$ increases, therefore the value that corresponds to $q_{crit}$, reported $\zeta_{\rm eff, min}$, is the minimum over all stable mass ratios. 

From Table~\ref{table} it is evident that any attempt to parametrize the adiabatic response in terms of the core mass alone will prove unsuccessful. The dependence of the donor's response on its radius has been noted in adiabatic models as well (Ge et al, in prep), and suggests that the correct response of giant branch donors can only be properly computed from detailed stellar evolutionary models. At the same time, it is difficult to generalize from the result of any such model an absolute criterion for MT, as it is difficult to determine when a true run-away occurs (see above). In principle the values for $q_{\rm{crit}}$ in Table 1 represent only minimum values for the given value of C used in eq. \ref{mdot}. 

\begin{table}[ht]
\begin{center}
\caption{Effective critical mass ratio for stable MT}
\begin{tabular}{c c c c c c c}
\hline
Donor & & & H-W & & Effective &\\
\hline
$M_{\mathrm{d}}$ $[M_{\odot}]$ & $R_{\mathrm{d}}$ $[R_{\odot}]$ & $m_{\mathrm{c}}/M_{\mathrm{d}}$ & $\zeta _{\mathrm{ad}}$ & $q_{\mathrm{crit}}$ & $\zeta_{\mathrm{eff,min}}$ &  $q_{\mathrm{crit}}$ \\
\hline
1  & 30  & 0.312 &  0.158 & 0.861 & -0.04 & 1.19\\
1  & 75  & 0.380 &  0.284 & 0.919 & -0.19 & 1.11\\
1  & 110 & 0.420 &  0.368 & 0.959 & -0.20 & 1.04\\
5  & 50  & 0.171 & -0.065 & 0.758 & 0.33  & 1.47 \\
5  & 68  & 0.171 & -0.065 & 0.758 & -0.02 & 1.11 \\
5  & 85  & 0.171 & -0.065 & 0.758 & -0.18 & 1.02 \\
10 & 300 & 0.223 &  0.013 & 0.787 & -0.58 & 0.64 \\
10 & 350 & 0.233 &  0.013 & 0.796 & -0.35 & 0.85 \\
20 & 950 & 0.308 &  0.151 & 0.86  & -0.20 & 1.05 \\
20(18.7) & 950  & 0.308 & 0.165 &  0.86  & -0.19 & 1.18 \\
\hline
\end{tabular}\label{table}
\tablecomments{Comparison of the mass-radius response exponent assuming a condensed polytrope \citep{Hjellming87} vs. our effective response from evolutionary calculations, and the resulting maximum mass ratios for stable MT. $\zeta_{\mathrm{ad}}$ as found from the approximation given in \cite{Soberman97}. Effective values for $q_{\rm{crit}}$ and $\zeta_{\rm{eff, min}}$ found for C = 1.}
\end{center}
\end{table}

Greater stability is found for fully conservative ML -- 
 no wind losses and all material transferred via RL overflow is accreted.
Any non-conservation only improves MT stability through the RL response ($\zeta_{\rm RL}$) \citep{Kal96, Han02}. 
This is of most importance in binaries with a compact accretor, 
as MT there would in the initial stage quickly exceed the Eddington limit. 
If non-conservation of MT is taken into account, 
$q_{\rm crit}$ can be even higher; for a donor of $5~M_\odot$ and $R=50~R_\odot$ we find $q_{\rm{crit}}\sim 1.61$, assuming all mass lost from the system leaves with the angular momentum of the accretor in a bipolar outflow. 
In a similar case, polytropic models would predict that only systems with $q<1.12$ are stable \citep[calculated as in][]{Soberman97}. 
Accounting for stellar wind also improves stability further (see Table 1 for 20$M_{\odot}$ case, although note that at $20M_{\odot}$ the situation is further complicated by a much greater depth of the envelope being superadiabatic). Wind losses may also be substantially enhanced by the companion \citep{TE88}, rendering MT far more stable or even avoidable. \cite{Eggbook06} argues that observational evidence suggests post-CE binaries may only be formed from progenitors where $q \gtrsim 4$.
Even for those systems in which run-away MT (and presumably a CE phase) 
appear to remain inevitable, at least 5--10$\%$ of the donor's mass may be transferred to the companion prior to such a phase. 
 
\cite{Ge_etal2010} note that their model for adiabatic ML must be understood 
with the caveat that realistically mass would only be lost through a small region 
in the vicinity of the inner Lagrangian point; fully 
3 dimensional calculations are needed in order to accurately describe the response of the donor.
Our studies obey similar limitations, but show the difference in the obtained stability 
of the MT within a similar framework.


\section{Conclusions}

While adiabatic ML models suggest that it is feasible for rapid ML to drive dramatic expansion of the envelope, this rests on the assumption that the superadiabatic surface can be removed completely. 

We conclude that: 
\begin{itemize}
\item thermal relaxation of the superadiabatic surface is reflected in the effective mass-radius response of the donor and affects the resulting stability strongly;

\item no parametrization can be done based only on $q$ and/or on the ratio of the core mass to the donor mass;

\item the response of the donor necessarily evolves with ML, rather than remaining static.
\end{itemize}

In principle one can only make strong conclusions for the response of any given donor on a case by case basis, using a detailed stellar evolutionary code. At the same time, one must acknowledge that attempting to determine $q_{\rm{crit}}$ for any given donor/core mass using such a code is inherently model-dependent.

Certainly ML from giant branch donors is more stable than previously thought from simplified adiabatic models. It is known that consideration of non-conservative MT shows stability is likely greater still, in particular for compact accretors in which $\dot M_{\rm{Edd}}$ is quickly reached.
However, to make any strong claims regarding the consequences for the formation rates of any given post-CE binary would be rather speculative without further population studies. 

\acknowledgements{
	We thank Stephen Justham for remarks on the manuscript. N.I. acknowledges support from NSERC Discovery and Canada Research Chair programs.}

\bibliographystyle{apj}
\bibliography{MT}

\end{document}